\documentclass[reqno]{amsart}

\usepackage[english]{babel}
\usepackage[utf8]{inputenc}
\usepackage[T1]{fontenc}
\usepackage{amsmath}
\usepackage{amssymb}
\usepackage{mathtools}
\usepackage{amsthm}
\usepackage{physics}
\usepackage{enumitem}

\usepackage[widespace]{fourier}

\usepackage{booktabs}
\usepackage{rotating}
\usepackage{tabularx}
\usepackage{multirow}

\usepackage{siunitx}

\usepackage{graphicx}
\usepackage{subfig}
\usepackage{wrapfig}
\usepackage{xcolor}
\usepackage{subfig}
\usepackage{tikz}
\usetikzlibrary{calc}
\usetikzlibrary{cd}

\usepackage{etoolbox} % for '\AtBeginEnvironment' macro
\AtBeginEnvironment{pmatrix}{\everymath{\displaystyle}}
\AtBeginEnvironment{bmatrix}{\everymath{\displaystyle}}

\DeclareMathOperator{\Id}{Id}
\DeclareMathOperator{\ud}{d}
\DeclareMathOperator{\Pic}{Pic}

\numberwithin{equation}{section}

\newcommand{\Pj}{\mathbb{P}}

\newcommand{\Z}{\mathbb{Z}}

\newcommand{\Cp}{\mathbb{C}}

\newcommand{\PcrossP}{\Pj^{1}\times\Pj^{1}}
\renewcommand{\epsilon}{\varepsilon}
\renewcommand{\imath}{\mathrm{i}}

\allowdisplaybreaks
\makeatletter 
\renewcommand{\pdv}[2]{\begingroup 
\@tempswafalse\toks@={}\count@=\z@ 
\@for\next:=#2\do 
{\expandafter\check@var\next\@nil
 \advance\count@\der@exp 
 \if@tempswa 
   \toks@=\expandafter{\the\toks@\,}% 
 \else 
   \@tempswatrue 
 \fi 
 \toks@=\expandafter{\the\expandafter\toks@\expandafter\partial\der@var}}% 
\frac{\partial\ifnum\count@=\@ne\else^{\number\count@}\fi#1}{\the\toks@}% 
\endgroup} 
\def\check@var{\@ifstar{\mult@var}{\one@var}} 
\def\mult@var#1#2\@nil{\def\der@var{#2^{#1}}\def\der@exp{#1}} 
\def\one@var#1\@nil{\def\der@var{#1}\chardef\der@exp\@ne} 
\makeatother

\theoremstyle{plain}
\newtheorem{theorem}{Theorem}[section]

\theoremstyle{definition}
\newtheorem{definition}{Definition}[section]

\theoremstyle{remark}
\newtheorem{remark}{Remark}[section]
\theoremstyle{remark}

\title{Space of initial values of a map with a quartic invariant}
\author{G. Gubbiotti \and N. Joshi}
\address{School  of  Mathematics  and  Statistics  F07,  The  University  of  Sydney,  NSW  2006, Australia}
\email{giorgio.gubbiotti@sydney.edu.au}
\email{nalini.joshi@sydney.edu.au}

\begin{document}

\begin{abstract}
    We compactify and regularize the space of initial values
    of a planar map with a quartic invariant and use this construction
    to prove its integrability in the sense of algebraic entropy. 
    The system turns out to have certain unusual properties, 
    including a sequence of points of indeterminacy in $\mathbb P^1\cross \mathbb P^1$.
    These indeterminacy points are shown to lie on a singular fibre of the mapping
    to a corresponding QRT system and provide the existence of a one-parameter 
    family of special solutions.
\end{abstract}

%% - subject classification and keywords
%% 2010 American Mathematical Society Subject Classification
%% Provide only ONE primary classification
%\classification{14E05; 14E15; 14H70; 14J17; 37F10;  39A10.}
%% Four or five keywords or phrases

\maketitle
\section{Introduction}

In this paper, we consider the birational maps
\begin{subequations}
    \begin{align}
        \varphi &\colon 
        \left( x,y \right) \mapsto
        \left( \frac{x (3 h y-1)}{4 h^2 x^2+2 h^2 y^2+h y-1}, 
        \frac{4 h x^2-h y^2-y}{4 h^2 x^2+2 h^2 y^2+h y-1} \right),
        \\
        \psi &\colon 
        \left( x,y \right) \mapsto
        \left( -\frac{x (3 h y+1)}{4 h^2 x^2+2 h^2 y^2-h y-1},
        -\frac{4 h x^2-h y^2+y}{4 h^2 x^2+2 h^2 y^2-h y-1} \right),
    \end{align}
    \label{eq:phipsi}
  \end{subequations}
  with $h\not=0$, whose actions preserve the following rational function
  \begin{equation}
    H\left( x, y\right) =
    \frac{x^{2} \left( x^{2}-y^{2} \right)}{%
    \left( 1-h^{2}y^{2} \right)\left( 1-8h^{2}x^{2}-h^{2}y^{2} \right)}\,,
  \label{eq:Hh}
\end{equation}
in the sense that $H\left( \varphi\left( x,y \right) \right)=H\left( x,y \right)$,
and $H\left( \psi\left( x,y \right) \right)=H\left( x,y \right)$. 
The maps $\varphi$ and $\psi$ are inverses in $\PcrossP$, i.e., we have
$\varphi\circ\psi=\psi\circ\varphi=\Id$. Their iteration leads to a dynamical system, whose geometric properties are considered in this paper. 
For simplicity, we sometimes refer to one of them, $\varphi$, as \textit{the map} and describe $H$ as an \textit{invariant} of the map. 

The maps \eqref{eq:phipsi} preserve the following
volume form \cite{CelledoniMcLachlanMcLarenOwrenQuispel2017}:
\begin{equation}
    \omega\left( x,y \right) = 
    \frac{\ud x \wedge \ud y}{x\left( x^{2}-y^{2} \right)} \implies 
    \omega\left( \varphi\left( x,y \right) \right)=\omega\left( \psi\left( x,y \right) \right)=
    \omega\left( x,y \right).
    \label{eq:measure}
  \end{equation}
Due to this measure-preserving property, the possession of the invariant $H$
suggests that $\varphi$ is \emph{integrable} \cite{HietarintaJoshiNijhoff2016}.
However, almost all the known integrable maps in the plane have biquadratic invariants, 
while the invariant $H(x,y)$ is evidently quartic.

The maps $\varphi$ and $\psi$ are clearly undetermined at certain points; e.g., $\varphi$ becomes $0/0$ as $(x, y)\to (0, -1/h)$. We blow up $\PcrossP$ at such points and consider the lifted maps on the resulting surface, called the \emph{initial value space}.
We show that the initial value space of the maps \eqref{eq:phipsi} possesses geometric properties that are not usually seen in other known integrable systems.  Another motivation of our study was the reconciliation of the unusual characteristics of this example with the observation that this map is transformable (via a birational mapping) to a QRT map, which has a biquadratic invariant \cite{VanDerKampCelledoniMcLachlanMcLarenOwrenQuispel2019}.

To state our results, we first recall that the degree $\deg(\varphi)$ of a rational map $\varphi:(x,y)\mapsto P(x,y)/Q(x,y)$ is given by $\max\bigl(\deg(P), \deg(Q)\bigr)$. Here, the degree of a polynomial of several variables is defined to be the sum of individual degrees in each variable, i.e., $\deg(P)=\deg_x(P)+\deg_y(P)$. A crucial concept in the study of the dynamics of the map is the degree of the $n$-th iterate: $d_n=\deg(\varphi^{(n)})$, as $n\to\infty$. The algebraic entropy of the map is defined by $\lim_{n\to\infty}\log(d_n)/n$.

The main new result of this paper is an algebro-geometric proof that the maps \eqref{eq:phipsi} possess quadratic degree growth \cite{Sakai2001,Takenawa2001,Takenawa2001JPhyA}. Consequently, its algebraic entropy vanishes. The proof is carried out by constructing the space of initial values,
on which the maps are not automorphisms but become analytically stable (see Definition \ref{def:as}). The surface contains an infinite sequence of indeterminacy points in $\mathbb P^1\cross \mathbb P^1$, which are spurious \cite{DillerFavre2001}, 
as their blow-up is not necessary to compute the growth of the map,
but are connected to the existence of a special one-parameter family of solutions.
To the best of our knowledge, this is the first non-linearsable
example where an infinite number of blow-ups appears to be needed
\cite{Takenawaatel2003,HayHoweseNakazonoShi2015}.

\subsection{Main result}

We refer the reader to terminology defined in Section \ref{sec:resolution}. The main result of this paper is summarised in the following theorem:
\begin{theorem}
    The following statements hold true:
    \begin{enumerate}[label={{\rm \thetheorem .(\alph *)}}, leftmargin=*]
    \item There exists an algebraic surface $S$, obtained by a composition of blow-ups $\pi:S \rightarrow \PcrossP$,
            on which the maps \eqref{eq:phipsi} are lifted to
            \emph{analytically stable maps}
            $\varphi^{*},\psi^{*}\colon S\to S$.
            \label{thm2:a}
        \item There exists a infinite sequence of points
            $\{ \theta_{k}\}_{k\in\Z\setminus\left\{ 0 \right\}}$
            generated by blowing up two indeterminacy points $\theta_{\pm1}$,
            corresponding to the ill-posed initial values of a one-parameter family 
            of solutions with initial conditions 
            $\left( x_{0},y_{0} \right)=\left( 0,a \right)$.
            \label{thm1:add}
        \item The degree of the $N$-th iterate of the maps \eqref{eq:phipsi}  given by:
            \begin{equation}
                d_{N} = \frac{2}{3}N^2
                -\frac{2}{9}\cos\left(\frac{2\pi N}{3}\right)
                +\frac{11}{9}.
                \label{eq:degreethm}
            \end{equation}
            \label{thm2:c}
        \item The invariant $H$ \eqref{eq:Hh} corresponds to the lowest-degree
            non-trivial element of the eigenspace relative to the
            eigenvalue $1$ of either $\varphi^{*}$ or $\psi^{*}$.
            \label{thm2:d}
    \end{enumerate}
    \label{thm:resolution2}
\end{theorem}

The proof of this theorem is presented in
section \ref{sec:resolution}.
The peculiarity of such an example is that the two lifted maps $\varphi^{*}$
and $\psi^{*}$ are not automorphisms.
This occurred in other examples of maps with linear growth
\cite{Takenawaatel2003,HayHoweseNakazonoShi2015},
but, to the best of our knowledge, this is the first time
that this occurrence appears in a map with quadratic growth.
We mention that the notion of analytical stability
was used to compute the growth of integrable maps in four-dimensions
\cite{CarsteaTakenawa2019JPhysA}.

\subsection{Background}
The geometric theory of discrete integrable systems 
has a long history (see \cite{DillerFavre2001} for a summary).
Its relationship with elliptic curves is the foundation of the study of 
QRT maps \cite{QRT1988,QRT1989,Tsuda2004}.
It also underlies the geometric formulation of 
discrete Painlev\'e equations \cite{Sakai2001}, which are expressed through Cremona transformations of a regular algebraic surface called the \emph{space of initial values},
 as the discrete equivalent of the Okamoto's 
description of the continuous Painlev\'e equations \cite{Okamoto1979}.

In recent years a procedure called 
\emph{Kahan-Hirota-Kimura (KHK) discretisation} became a popular way 
of  producing integrable discrete equations from systems of integrables ODEs.
This procedure was presented first by W. Kahan in a series of unpublished
lecture notes \cite{Kahan1993} as a method to obtain better numerical approximations.
The interest of the integrable systems community on this procedure
arose after this method was proved to produce an integrable discretisation of
the Lagrange top \cite{HirotaKimura2000}, see
\cite{PetreraPfadlerSuris2009,PetreraSuris2010,PetreraPfadlerSuris2011,
CelledoniMcLachlanOwrenQuispel2013,CelledoniMcLachlanOwrenQuispel2014}.

From the results of \cite{CelledoniMcLachlanOwrenQuispel2014}, it follows
that the KHK discretisation of a two-dimensional system with a cubic
Hamiltonian is always integrable.
This observation led to the consideration of non-standard
quadratic systems in \cite{CelledoniMcLachlanMcLarenOwrenQuispel2017},
whose KHK discretisation possesses a higher-degree polynomial invariant.
In this paper, we consider the KHK discretisation of one of the
two systems discovered independently in
\cite{CelledoniMcLachlanMcLarenOwrenQuispel2017,PetreraZander2017}.

The maps \eqref{eq:phipsi} form a discretisation of the octahedral reduced Nahm system \cite{Hitchinetal1995,PetreraPfadlerSuris2011}. We note that a broader class of maps containing \eqref{eq:phipsi}
was presented in \cite{VanDerKampCelledoniMcLachlanMcLarenOwrenQuispel2019}, 
where it was shown that the invariant \eqref{eq:Hh} can be reduced
to an invariant of QRT type, dropping its total degree by four through
a fractional linear transformation. 
However, the corresponding pencil factors to give a QRT pencil times a  multiple
singular fibre given by $x^{4}=0$.

\subsection{Outline of the paper}
The plan of the paper is as follows. The transformation of the maps \eqref{eq:phipsi} to a QRT system is given in section 
\ref{sec:background}.
Section \ref{sec:resolution} is devoted to the proof of Theorem 
\ref{thm:resolution2}.
Finally, in section \ref{sec:conclusions}, 
we summarize our results and discuss open questions. 

\section{The relation to a QRT system}
\label{sec:background}
In this section, we recall the relation between the maps \eqref{eq:phipsi} and QRT systems \cite{QRT1988,QRT1989} provided by \cite{VanDerKampCelledoniMcLachlanMcLarenOwrenQuispel2019}.

The proof in \cite{VanDerKampCelledoniMcLachlanMcLarenOwrenQuispel2019} 
can be summarised as follows. Consider the change of variables:
\begin{equation}
    x =\frac{2}{u+v},
    \quad
    y = -\,\frac{u-v}{h(u+v)}.
    \label{eq:chtoqrt}
\end{equation}
Then the invariant \eqref{eq:Hh} is mapped to:
\begin{equation}
    K(h) = \frac{1}{4h^2} \frac{(u-v-2h) (u-v+2h)}{uv(8 h^2-u v)},
    \label{eq:Kh}
\end{equation}
which is the ratio of two symmetric biquadratic polynomials,
hence an invariant of QRT type.
The fact that the quartic invariant \eqref{eq:Hh}
reduces to the biquadratic invariant \eqref{eq:Kh} through
the fractional linear transformation \eqref{eq:chtoqrt}
means that some factorisation is happening. We will show
below, that this phenomenon appears clearly if instead of 
the rational invariants we consider the associated 
\emph{covariant pencils of curves}.

Under the transformation \eqref{eq:chtoqrt}
the maps \eqref{eq:phipsi} become:
\begin{equation}
    \widetilde{\varphi} \colon 
    \left( u,v \right) \mapsto
    \left( v, -\frac{8 h^2-u v}{2 u-v}\right),
    \quad
    \widetilde{\psi} \colon 
    \left( u,v \right) \mapsto
    \left( \frac{8 h^2-u v}{u-2 v} ,u \right).
    \label{eq:phipsit}
\end{equation}
The maps \eqref{eq:phipsit} are symmetric QRT maps and define a
second-order recurrence relation.

The above results were presented in \cite{VanDerKampCelledoniMcLachlanMcLarenOwrenQuispel2019}.
Now we consider the reduction from a quartic to
a biquadratic invariant.
Consider the pencil of curves associated to \eqref{eq:Hh}:
\begin{equation}
    p\left( x,y \right) =
    x^{2} \left( x^{2}-y^{2} \right)+
    \varepsilon_{0}\left( 1-h^{2}y^{2} \right)\left( 1-8h^{2}x^{2}-h^{2}y^{2} \right).
    \label{eq:quartic}
\end{equation}
Here $\varepsilon_{0}\in\Pj^{1}$ is the parameter of the pencil.
Consider now the pencil associated to the invariant 
\eqref{eq:Kh}:
\begin{equation}
    q\left( u,v \right) = (u-v-2h) (u-v+2h) + 4\varepsilon_{0}h^2 uv(8 h^2-u v).
    \label{eq:biquadratic}
\end{equation}
Then applying the coordinate transformation \eqref{eq:chtoqrt}
we obtain:
\begin{equation}
    p\left( u,v \right)
    = -\frac{4 q\left( u,v \right)}{h^{2}\left( u+v \right)^{4}}
    = -\frac{x^{4}}{4h^{2}}q\left( u,v \right).
    \label{eq:equiv}
\end{equation}
Therefore, the two pencils are equivalent except on the  one-dimensional
\emph{singular multiple fibre $x=0$}.
This is what causes the bi-degree to drop from $\left( 4,4 \right)$ of $p$ to
$\left( 2,2 \right)$ of $q$.

\section{Space of initial values of maps \eqref{eq:phipsi}}
\label{sec:resolution}

In this section, we prove Theorem \ref{thm:resolution2}.
We start by constructing a rational surface $S$, which is obtained by blowing up  a sequence of points of indeterminacy $p_i$, through monoidal transformations  $\pi_i:S_i\to S_{i-1}$ with centre $p_i$, where $S_0=\PcrossP$. The set $S_i$ is called the strict transform of $S_{i-1}$ and the map $\varphi$ is said to be \emph{lifted} to $S_i$. Each $\pi_i$ induces an isomorphism of $S_i-\pi_i^{-1}(p_i)$ onto $S_{i-1}-p_i$ \cite[\S V.3]{hartshorne2013}. Here the set $\pi_i^{-1}(p_i)$ is isomorphic to $\mathbb P$ and is called an \emph{exceptional line}. The composition of all monoidal transformations $\pi_i$ will be denoted by $\pi\colon S \to \PcrossP$.

The following definition is equivalent to the standard definition of analytical stability given in \cite{DillerFavre2001,Takenawaatel2003}. (We note that the equivalency is stated as a remark in \cite[\S 3.3]{Takenawaatel2003}.)
\begin{definition}[\!\cite{Takenawaatel2003}]\label{def:as}
    Suppose we are given a map $\varphi\colon\PcrossP\to\PcrossP$, lifted to an initial value space $S$, by a sequence of blow-ups $\pi:S \to \PcrossP$.
    If the lifted map $\varphi^{*}\colon S \to S$ does not give rise to any iteration sequence of the form:
    \begin{equation}
        \mathcal{E} \xrightarrow{\varphi^{*}} p_{1} \xrightarrow{\varphi^{*}}
        p_{2}\xrightarrow{\varphi^{*}} \dots \xrightarrow{\varphi^{*}}
        p_{K} \xrightarrow{\varphi^{*}} \mathcal{E}',
        \label{eq:singpatt}
    \end{equation}
    where $\mathcal{E}$, $\mathcal{E}'$ are one-dimensional sub-varieties 
    and $p_{1}$, $p_{2}$, \dots, $p_{K}$ is a finite set of points.
    Then the map $\varphi^{*}$ is called \emph{analytically stable}. 
    \label{def:anstable}
\end{definition}
\noindent Heuristically an analytically stable map is a map which does not possess any \emph{periodic singularity pattern}.

Let the affine charts of $\PcrossP$ be given by $\left( x,y \right)$,
$\left( 1/x,y \right)$, $\left( x,1/y\right)$, and $\left( 1/x,1/y \right)$,
where the lines at infinity correspond to $1/x=0$ and $1/y=0$ respectively.
Suppose $(x_0, y_0)$ is a point of indeterminacy of $\varphi$. Its blow up is given by
\begin{equation}
    \left( x,y \right) \leftarrow
    \left( x-x_{0},\frac{y-y_{0}}{x-x_{0}} \right)
    \bigcup
    \left( \frac{x-x_{0}}{y-y_{0}},y-y_{0} \right).
    \label{eq:blowup}
\end{equation}

The map $\varphi$ is indeterminate at the points
$\left(0, -1/h\right)$, 
$\left(0, 1/2h\right)$, 
$\left(1/3h,1/3h\right)$,
$\left(-1/3h,1/3h\right)$
and $\left(\infty, \infty\right)$.
We obtain the following sequence of blow-ups. (For simplicity,
we state only one coordinate chart at each step of \eqref{eq:blowup} and indicate the centre and exceptional line of each blow-up above and below the arrow corresponding to the monoidal transformation.)
\begin{subequations}
    \begin{gather}
        \left( x,y \right) \xleftarrow[E_{1}]{\left( 0,-1/h \right)}
        \left( x,\frac{y+1/h}{x} \right)
        \xleftarrow[E_{2}]{\left( 0,0 \right)}
        \left( x,\frac{y+1/h}{x^{2}} \right),
        \label{eq:bpphi1}
        \\
        \left( x,y \right) \xleftarrow[E_{3}]{\left( 1/3h,-1/3h \right)}
        \left( x-\frac{1}{3h},\frac{y+1/3h}{x-1/3h} \right)
        \label{eq:bppsi3}
        \\
        \left( x,y \right) \xleftarrow[E_{4}]{\left( -1/3h,-1/3h \right)}
        \left( x+\frac{1}{3h},\frac{y+1/3h}{x+1/3h} \right)
        \label{eq:bppsi4}
        \\
        \left( x,y \right) \xleftarrow[E_{5}]{\left( \infty,\infty \right)}
        \left( \frac{1}{x},\frac{x}{y} \right).
        \label{eq:bpphi5}
    \end{gather}
    \label{eq:bpphi}
\end{subequations}
We will consider the point $\left( 0,1/2h \right)$ separately later.
The images of the exceptional lines $E_{4}$ and $E_{13}$ 
under $\varphi$ are given by
\begin{equation}
    \varphi^{*}\left(E_{3}\right)= \left(\frac{1}{h},-\frac{1}{h}  \right),
    \quad
    \varphi^{*}\left(E_{4}\right)= \left(-\frac{1}{h},\frac{1}{h}  \right),
    \label{eq:imb34}
  \end{equation}
  indicating additional points of indeterminacy.
  
This implies that we need to blow-up the corresponding image points:
\begin{subequations}
    \begin{gather}
        \left( x,y \right) \xleftarrow[E_{6}]{\left( 1/h,-1/h \right)}
        \left( x-\frac{1}{h},\frac{y+1/h}{x-1/h} \right),
        \label{eq:bpphi6}
        \\
        \left( x,y \right) \xleftarrow[E_{7}]{\left( -1/h,-1/h \right)}
        \left( x+\frac{1}{h},\frac{y+1/h}{x+1/h} \right).
        \label{eq:bpphi7}
    \end{gather}
    \label{eq:bpphibis}
\end{subequations}
The images of the exceptional lines $E_{6}$ and $E_{7}$ under 
$\varphi$ are:
\begin{equation}
    \varphi^{*}\left(E_{6}\right)= \left(-\frac{1}{h},\frac{1}{h}  \right),
    \quad
    \varphi^{*}\left(E_{7}\right)= \left(\frac{1}{h},\frac{1}{h}  \right).
    \label{eq:imb67}
  \end{equation}
  
Again, we blow-up the corresponding image points:
\begin{subequations}
    \begin{gather}
        \left( x,y \right) \xleftarrow[E_{8}]{\left( 1/h,1/h \right)}
        \left( x-\frac{1}{h},\frac{y-1/h}{x-1/h} \right),
        \label{eq:bpphi8}
        \\
        \left( x,y \right) \xleftarrow[E_{9}]{\left( -1/h,1/h \right)}
        \left( x+\frac{1}{h},\frac{y-1/h}{x+1/h} \right).
        \label{eq:bpphi9}
    \end{gather}
    \label{eq:bpphitris}
\end{subequations}
The images of the exceptional lines $E_{8}$ and $E_{9}$ are now one-dimensional curves. This implies that the lifted map $\varphi^{*}$ obtained by blowing-up all the 
the above points satisfies Definition \ref{def:anstable} and is analytically stable. 

Now we consider the map $\psi$, which is indeterminate at the points
$\left(0,1/h\right)$, 
$\left(0, -1/2h\right)$, 
$\left(1/3h,-1/3h\right)$ 
and $\left( -1/3h,-1/3h\right)$.
The two maps share the indeterminate point at $\left( \infty,\infty \right)$,
but it is automatically resolved under the transformation \eqref{eq:bpphi5}.
As above, we obtain the following sequence of blow-ups:
\begin{subequations}
    \begin{gather}
        \left( x,y \right) \xleftarrow[E_{10}]{\left( 0,1/h \right)}
        \left( x,\frac{y-1/h}{x} \right)
        \xleftarrow[E_{11}]{\left( 0,0 \right)}
        \left(  x,\frac{y-1/h}{x^{2}}\right)
        \label{eq:bppsi1}
        \\
        \left( x,y \right) \xleftarrow[E_{12}]{\left( 1/3h,1/3h \right)}
        \left( x-\frac{1}{3h},\frac{y-1/3h}{x-1/3h} \right),
        \label{eq:bpphi3}
        \\
        \left( x,y \right) \xleftarrow[E_{13}]{\left( -1/3h,1/3h \right)}
        \left( x+\frac{1}{3h},\frac{y-1/3h}{x+1/3h} \right).
        \label{eq:bpphi4}
    \end{gather}
    \label{eq:bppsi}
\end{subequations}
We will consider the point $\left( 0,-1/2h \right)$ later.
The images of the resulting exceptional lines are one-dimensional, showing that
 the lifted map $\psi^{*}$ after blowing-up all
these points is analytically stable (see Definition 
\ref{def:anstable}).

\begin{remark}\label{rem:ivs}
 Consider the set of indeterminacy of a map $\varphi$ defined by
\[
\mathcal I(\varphi)=\left\{p\in \PcrossP : \varphi \ \textrm{is indeterminate at}\ p\right\}.
\]
A space $\mathcal S$ is called a \emph{space of initial values} when the indeterminacy set of the lift of $\varphi$ to $\mathcal S$ is empty  \cite{Okamoto1977,Sakai2001,Takenawa2001}. We denote the space $\PcrossP$ blown up at points given in Equations \eqref{eq:bpphi}, \eqref{eq:bpphibis}, \eqref{eq:bpphitris}, \eqref{eq:bppsi} by $S$, and use it as a basis for our arguments below. We will see below that $S$ is not strictly a space of initial values. However, for our purposes, it is sufficient to consider the maps $\varphi$, $\psi$ on $S$ where they are analytically stable. 
\end{remark}

The only remaining points to consider are  $\theta_{\pm1}:=\left( 0,\pm1/2h \right)$. 
Blowing up $\theta_{\pm1}$ gives
\begin{equation}
    \left( x,y \right) \xleftarrow[\Theta_{\pm1}]{\theta_{\pm}}
    \left( x,\frac{y\mp1/2h}{x} \right),
    \label{eq:specialbp}
  \end{equation}
  where $\Theta_{\pm1}$ are the corresponding exceptional lines.
%\begin{subequations}
%    \begin{align} 
%        \left( x,y \right) \xleftarrow[\Theta_{1}]{\left( 0,1/2h \right)}
%        \left( x,\frac{y-1/2h}{x} \right),
%        \label{eq:bpphi2}
%        \\
%        \left( x,y \right) \xleftarrow[\Theta_{-1}]{\left( 0,-1/2h \right)}
%        \left( x,\frac{y+1/2h}{x} \right)
%        \label{eq:bppsi2}
%    \end{align}
%    \label{eq:specialbp}
%\end{subequations}
The images of these exceptional lines are:
\begin{equation}
    \psi^{*} \left( \Theta_{1}  \right) =
    \left( 0, \frac{1}{4h} \right)=:\theta_{2},
    \quad 
    \varphi^{*} \left( \Theta_{-1}  \right) =
    \left( 0, -\frac{1}{4h} \right)=:\theta_{-2}.
    \label{eq:theta2m2birth}
\end{equation}
Blowing up these two points $\theta_{\pm2}$ leads to exceptional lines whose images are points again. We can show inductively that iterating this process leads to an infinite sequence of points $\theta_{k} := \left( 0,1/(2kh) \right)$, 
$k\in\Z\setminus\left\{ 0 \right\}$, with
\begin{equation}
    \left( x,y \right) 
    \xleftarrow[\Theta_{k}]{\theta_{k}}
    \left( x,\frac{y-1/2hk}{x} \right),
    \label{eq:thetakbp}
\end{equation}
for which the following relations hold:
\begin{equation}
    \varphi^{*}\left( \Theta_{k} \right) = \theta_{k-1},
    \quad
    \psi^{*}\left( \Theta_{k} \right) = \theta_{k+1},
    \quad
    k\in\Z\setminus\left\{ \pm1,0 \right\}.
    \label{eq:thetakpbirth}
\end{equation}
This implies the existence of infinite sequences of images under $\varphi^{*}$ and $\psi^{*}$:
\begin{subequations}
    \begin{gather}
        \Theta_{-1}  \xrightarrow{\varphi^{*}} \theta_{-2} \xrightarrow{\varphi^{*}}
        \theta_{-3}\xrightarrow{\varphi^{*}}\theta_{-4} \xrightarrow{\varphi^{*}}\theta_{-5} \xrightarrow{\varphi^{*}}\cdots, 
        \\
        \Theta_{1}  \xrightarrow{\psi^{*}} \theta_{2} \xrightarrow{\psi^{*}}
        \theta_{3}\xrightarrow{\psi^{*}}\theta_{4} \xrightarrow{\psi^{*}}\theta_{5} \xrightarrow{\psi^{*}}\cdots,
    \end{gather}
    \label{eq:thkseq}
\end{subequations}
which never terminates on a one-dimensional sub-variety. 
This violates the requirement of analytical stability.

\begin{figure}[hbt]
    \centering
    \begin{tikzpicture}
        \def\xmin{3};
        \def\xmed{3};
        \def\xmax{7};
        \def\hvar{0.45};
        \draw[dashed] (-\xmin,0)--(\xmed+\xmin/2,0);
        \draw[dashed] (-\xmin,\xmax-\xmin)--(\xmed,\xmax-\xmin);
        \draw[dashed] (-\xmin+\xmin,-\xmin)--(-\xmin+\xmin,\xmed+\xmin/2);
        \draw[dashed] (\xmax-\xmin,-\xmin)--(\xmax-\xmin,\xmed);
        \coordinate(l1) at (0,1/\hvar);
        \coordinate(l2) at (0,-1/\hvar);
        \coordinate(thetam1) at (0,-1/\hvar/2);
        \coordinate(theta1) at (0,1/\hvar/2);
        \coordinate(b1) at (1/\hvar/3,1/\hvar/3);
        \coordinate(b2) at (1/\hvar/3,-1/\hvar/3);
        \coordinate(b3) at (-1/\hvar/3,-1/\hvar/3);
        \coordinate(b4) at (-1/\hvar/3,1/\hvar/3);
        \coordinate(b5) at (1/\hvar,1/\hvar);
        \coordinate(b6) at (1/\hvar ,-1/\hvar );
        \coordinate(b7) at (-1/\hvar ,-1/\hvar );
        \coordinate(b8) at (-1/\hvar ,1/\hvar );
        \coordinate(rp1) at (-\xmin/2,1/\hvar/2);
        \coordinate(rp2) at (\xmin/2,1/\hvar/2);
        \coordinate(rm1) at (-\xmin/2,-1/\hvar/2);
        \coordinate(rm2) at (\xmin/2,-1/\hvar/2);
        \draw[thick] (\xmed+\xmin/2,\xmed-\xmin/2)--node[right] {\small $E_{5}$} (\xmed-\xmin/2,\xmed+\xmin/2);
        \draw[thick] (-\xmin/3,1/\hvar)--(l1)-- (\xmin/3,1/\hvar) node[anchor=south] {\small $E_{10}-E_{11}$};
        \draw[thick] (-\xmin/4,1/\hvar-\xmin/4)--node[above left] {\small $E_{11}$} (-\xmin/4,1/\hvar+\xmin/4);
        \draw[thick] (-\xmin/3,-1/\hvar)--(l2)-- (\xmin/3,-1/\hvar) node[anchor=north] {\small $E_{1}-E_{2}$};
        \draw[thick] (-\xmin/4,-1/\hvar-\xmin/4)--node[below left] {\small $E_{2}$} (-\xmin/4,-1/\hvar+\xmin/4);
        \draw[thin] ($(b7)-(\hvar,\hvar)$) -- ($(b5)+(\hvar,\hvar)$);
        \draw[thin] ($(b8)-(\hvar,-\hvar)$) -- ($(b6)+(\hvar,-\hvar)$);
        \draw[thick] ($(b1)-(\xmin/6,-\xmin/6)$)--node[right] {\small $E_{12}$} ($(b1)+(\xmin/6,-\xmin/6)$);
        \draw[thick] ($(b2)-(\xmin/6,\xmin/6)$)--node[right] {\small $E_{3}$} ($(b2)+(\xmin/6,\xmin/6)$);
        \draw[thick] ($(b3)-(\xmin/6,-\xmin/6)$)--node[right] {\small $E_{4}$} ($(b3)+(\xmin/6,-\xmin/6)$);
        \draw[thick] ($(b4)-(\xmin/6,\xmin/6)$)--node[right] {\small $E_{13}$} ($(b4)+(\xmin/6,\xmin/6)$);
        \draw[thick] ($(b5)-(\xmin/12,-\xmin/12)$) -- ($(b5)+(\xmin/4,-\xmin/4)$) node[anchor=north west] {\small $E_{8}$};
        \draw[thick] ($(b6)-(\xmin/12,\xmin/12)$) -- ($(b6)+(\xmin/4,\xmin/4)$) node[anchor=south west] {\small $E_{6}$} ;
        \draw[thick] ($(b7)+(\xmin/12,-\xmin/12)$) -- ($(b7)-(\xmin/4,-\xmin/4)$) node[anchor=south east] {\small $E_{7}$};
        \draw[thick] ($(b8)+(\xmin/12,\xmin/12)$) -- ($(b8)-(\xmin/4,\xmin/4)$) node[anchor=north east] {\small $E_{9}$};
    \end{tikzpicture}
    \caption{The figure denotes the surface $S$ described in Remark \ref{rem:ivs}. The dashed lines are coordinate axes (in the finite and infinite charts). The exceptional lines $E_j$, $j=1,\ldots, 13$,
      are defined in Equations \eqref{eq:bpphi},
      \eqref{eq:bpphibis}, \eqref{eq:bpphitris}, \eqref{eq:bppsi}. 
    }\label{fig:surf}
\end{figure}
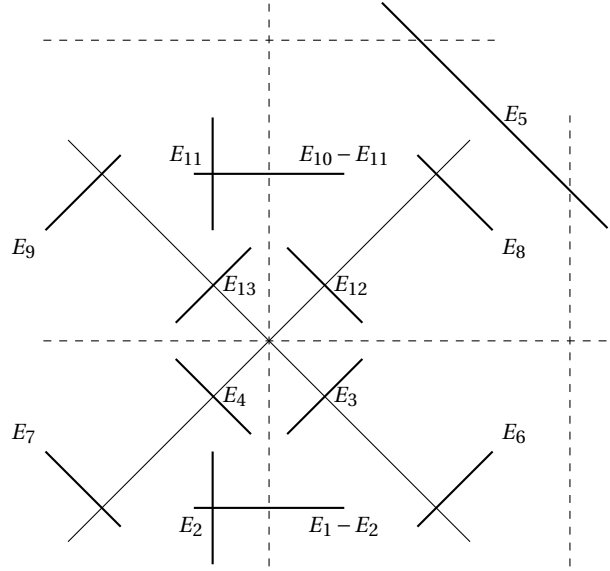

In conclusion, the maps $\varphi^{*}$ and $\psi^{*}$, are analytically stable on $S$ defined in Remark \ref{rem:ivs}. 
This proves the first  statement \ref{thm2:a} of Theorem \ref{thm:resolution2}.
A schematic representation of the surface $S$ obtained after 
performing all the blowups is given in Figure \ref{fig:surf}.

Now we prove the claim \ref{thm1:add} of Theorem \ref{thm:resolution2}.
Consider the orbit of the the one-parameter family of initial points $\left( x_{0},y_{0} \right)=\left( 0,a \right)$:
\begin{subequations}
    \begin{gather}
        \begin{aligned}
            \varphi(0,a) &= \left( 0,\frac{a}{1-2ah} \right),
            \quad 
            \varphi^{2}(0,a) = \left( 0,\frac{a}{1-4ah} \right),
            \\
            \varphi^{3}(0,a) &= \left( 0,\frac{a}{1-6ah} \right),
            \quad
            \varphi^{4}(0,a) = \left( 0,\frac{a}{1-8ah} \right)
            \dots,
        \end{aligned}
        \label{eq:orbyphi}
        \\
        \begin{aligned}
            \psi(0,a) &= \left( 0,\frac{a}{1+2ah} \right),
            \quad 
            \psi^{2}(0,a) = \left( 0,\frac{a}{1+4ah} \right),
            \\
            \psi^{3}(0,a) &= \left( 0,\frac{a}{1+6ah} \right),
            \quad
            \psi^{4}(0,a) = \left( 0,\frac{a}{1+8ah} \right) 
            \dots.
        \end{aligned}
        \label{eq:orbypsi}
    \end{gather}
    \label{eq:orby}
\end{subequations}
By induction, it follows that the orbit of  
$\left( x_{0},y_{0} \right)=\left( 0,a \right)$ is given by
\begin{equation}
    \left\{ \left( 0,\frac{a}{1-2akh} \right) \right\}_{k\in\Z}.
    \label{eq:orbclosed}
\end{equation}
However, the sequence \eqref{eq:orbclosed} is ill-defined on
the points $\left( 0,1/2kh \right)$, $k\in\Z\setminus\left\{ 0 \right\}$. 
Hitting one of this points will result in the orbit falling into
one of the orbits given in equation \eqref{eq:thkseq}. This proves the statement \ref{thm1:add}.

Now consider the statement \ref{thm2:c} of Theorem \ref{thm:resolution2}.
The Picard lattice of the algebraic surface $\pi\colon S\to\PcrossP$
is given by the following 15-dimensional $\Z$-module:
\begin{equation}
    \Pic\left( S \right) =
    \Z H_{x} +\Z H_{y} +
    \sum_{i=1}^{13}\Z E_{i}.
    \label{eq:PicSigma}
\end{equation}
The maps $\varphi^{*}$ and $\psi^{*}$ have the following linear actions on $\Pic\left( S \right)$:
\begin{subequations}
    \begin{align}
        \varphi^{*}&\colon
        \begin{pmatrix}
            H_{x}
            \\
            H_{y}
            \\
            E_{1},
            \\
            E_{2},E_{3},E_{4}
            \\
            E_{5},
            \\
            E_{6},E_{7},E_{8},E_{9}
            \\
            E_{10},E_{11}
            \\
            E_{12},
            \\
            E_{13}
        \end{pmatrix}
        \to
        \begin{pmatrix}
            2 H_x+2 H_y-E_{3}-E_{4}-2 E_{5}-E_{10}
            \\
            2 H_x+2 H_y-E_{3}-E_{4}-2 E_{5}-E_{10}-E_{11}
            \\
            H_x+H_y-E_{3}-E_{4}-E_{5} 
            \\
            H_x-E_{5}, 
            E_{6},
            E_{7},
            \\
            2 H_x+2 H_y-E_{3}-E_{4}-2 E_{5}-E_{10}-E_{11}
            \\
            E_{9},
            E_{8},
            E_{12},
            E_{13}
            \\
            E_{1},
            E_{2}
            \\
            H_x+H_y-E_{3}-E_{5}-E_{10}
            \\
            H_x+H_y-E_{4}-E_{5}-E_{10}
        \end{pmatrix},
        \label{eq:phipic}
        \\
        \psi^{*}&\colon
        \begin{pmatrix}
            H_{x}
            \\
            H_{y}
            \\
            E_{1},E_{2},
            \\
            E_{3},
            \\
            E_{4}
            \\
            E_{5},
            \\
            E_{6},E_{7},E_{8},E_{9}
            \\
            E_{10},
            \\
            E_{11},E_{12},E_{13}
        \end{pmatrix}
        \to
        \begin{pmatrix}
            2 H_x+2 H_y-E_{1}-2 E_{5}-E_{12}-E_{13}
            \\
            2 H_x+2 H_y-E_{1}-E_{2}-2 E_{5}-E_{12}-E_{13}
            \\
            E_{10},
            E_{11}
            \\
            H_x+H_y-E_{1}-E_{5}-E_{12},
            \\
            H_x+H_y-E_{1}-E_{5}-E_{13},
            \\
            2 H_x+2 H_y-E_{1}-E_{2}-2 E_{5}-E_{12}-E_{13}
            \\
            E_{3},
            E_{4},
            E_{7},
            E_{6}
            \\
            H_x+H_y-E_{5}-E_{12}-E_{13},
            \\
            H_x-E_{5},
            E_{8},
            E_{9}
        \end{pmatrix}.
    \end{align}
    \label{eq:psipic}
\end{subequations}
Iterating of the map $\varphi^{*}$ we have:
\begin{equation}
    \left( \varphi^{*} \right)^{N}\left( H_{x} \right) = 
    \alpha_{N} H_{x}+ \alpha_{N} H_{y}+ \dots,
    \quad
    \left( \varphi^{*} \right)^{N}\left( H_{y} \right) = 
    \left(\alpha_{N}-\beta_{N}\right) H_{x}+ \alpha_{N} H_{y}+ \dots,
    \label{eq:MphiN}
\end{equation}
where, for each integer $N$, we have
\begin{equation}
        \alpha_{N} =\frac{2}{3}N^{2} +\frac{11}{9} - \frac{2}{9}\cos\left( \frac{2N\pi}{3} \right),
        \quad
        \beta_{N} = \frac{1}{3} + \frac{2}{3}\cos\left( \frac{2N\pi}{3} \right).
    \label{eq:seq}
\end{equation}

Takenawa \cite[\S 6]{Takenawa2001JPhyA} showed that the coefficients of $H_{x}$ and $H_{y}$ arising in the iterated image of a map provide a geometric estimate of the degree of the map. In particular, the degree $d_{N}$ of the $N$th iterate of $\varphi$ is given by the
maximum of these coefficients.
In our case, since $\beta_{N}\in\left\{ 0,1 \right\}$ for all integer $N$, 
we obtain the expression
\begin{equation}
    d_{N} \equiv \alpha_{N} = \frac{2}{3}N^2-\frac{2}{9}\cos\left(\frac{2\pi N}{3}\right)
    +\frac{11}{9},
    \label{eq:dnfin}
\end{equation}
for the degree of the $N$-th iterate of the map. This proves the formula \eqref{eq:degreethm}.
The same considerations apply for the map $\psi^{*}$, which also provides 
 the degree growth \eqref{eq:dnfin}. This proves statement \ref{thm2:c} of Theorem \ref{thm:resolution2}. Clearly $d_{N}\sim N^{2}$ as $N\to\infty$ and this shows that the
algebraic entropy of the maps \eqref{eq:phipsi} vanishes.  

Finally, we prove statement \ref{thm2:d} of Theorem \ref{thm:resolution2}.
To this end, we consider the eigenspace of the eigenvalue $\lambda=1$ 
of the map $\varphi^{*}$ or $\psi^*$ on the Picard lattice.
This eigenspace of $\varphi^{*}$ is generated by the following elements:
\begin{subequations}
    \begin{align}
        D_{1} &= H_{x}+H_{y}-E_{4}-E_{5}-E_{7}-E_{8}-E_{12},
        \\
        D_{2} &= H_{x}+H_{y}-E_{3}-E_{5}-E_{6}-E_{9}-E_{13},
        \\
        D_{3} &= H_{y}-E_{1}-E_{2} -E_{10}-E_{11},
        \\
        D_{4} &= E_{1}-E_{3}-E_{6}-E_{9}+E_{10}-E_{13}.
    \end{align}
    \label{eq:Db}
\end{subequations}
To find an invariant, we search for a linear combination
$D=n_{1}D_{1}+n_{2}D_{2}+n_{3}D_{3}+n_{4}D_{4}$ which is realised
by a curve in $\PcrossP$.
From a numerical search, we find that a suitable set of values is
$\left( n_{1},n_{2},n_{3},n_{4} \right)=\left( 1,3,0,-2 \right)$.
That is:
\begin{equation}
    D 
    \begin{aligned}[t]
        &= 
        4H_x + 4 H_y
        -2E_{1}-E_{3}-E_{4}-4 E_{5}
        \\
        &-E_{6}-E_{7}-E_{8}-E_{9}
        -2 E_{10}-E_{12}-E_{13}.
    \end{aligned}
    \label{eq:Dsol}
\end{equation}
The corresponding curve is given by a multiple of the quartic pencil 
\eqref{eq:quartic}.
A similar computation holds for the map $\psi^{*}$ and gives
the same invariant \eqref{eq:Dsol}. These serve to prove the statement \ref{thm2:d} and concludes our proof of
Theorem \ref{thm:resolution2}.

\section{Conclusion}
\label{sec:conclusions}

In this paper, we constructed an algebraic surface $S$ by resolving (or blowing up) a sequence of points in $\PcrossP$ given by singularities of the two maps $\varphi$ and $\psi$ \eqref{eq:phipsi}. We showed that the lifted maps are analytically stable on $S$, but are not automorphisms of $S$. 
We used this construction to prove that the growth of degrees in the 
orbits generated by the maps is quadratic in the step $N$ -- see formula \eqref{eq:degreethm} --
thus proving the integrability of the maps \eqref{eq:phipsi} in 
the sense of algebraic entropy \cite{BellonViallet1999,Takenawa2001JPhyA}. 

In this study, we encountered a spurious infinite sequence of exceptional lines 
$\left\{ \Theta_{k} \right\}_{k\in\Z\setminus\left\{ 0 \right\}}$ 
which are mapped to points under $\varphi$ and $\psi$.
We showed that this sequence gives two orbits, which never terminate
on a variety of codimension 1. However, the corresponding sequence of indeterminacy points
$\left\{ \theta_{k} \right\}_{k\in\Z\setminus\left\{ 0 \right\}}$
lies entirely on the singular fibre $\left\{ x=0 \right\}$, which plays a special role in the transformation from the pencil 
$p$ \eqref{eq:quartic}, covariant under \eqref{eq:phipsi}, 
to the pencil $q$ \eqref{eq:biquadratic}, covariant under the associated 
QRT map \eqref{eq:phipsit}. 
The quartic pencil $p$ factors into four factors of the vertical line 
$\{x=0\}$ and the biquadratic pencil $q$.
Although not needed to compute the algebraic entropy of the map, 
this sequence of indeterminate points turn out to correspond to 
inadmissible initial values characterising  a one-parameter family of solutions.

The occurrence of such solutions (in $\PcrossP$) show that spurious indeterminacy
points can appear also in integrable examples.
See \cite{HayHoweseNakazonoShi2015} for several examples of this
occurrence in the linearisable case.

While revising this paper, we became aware of the preprint \cite{Zander2020},
posted after our manuscript was posted on \texttt{arXiV},
which considered the integrability of the two-di\-men\-sion\-al version of 
the maps presented in \cite{PetreraZander2017} with the method
of \emph{singular orbits}, see e.g. \cite{BedfordKim2006}.
These include the example we presented in this paper,
with a different compactification of $\Cp^{2}$, namely $\Pj^{2}$.
The results presented in \cite{Zander2020} agree with the ones in
our paper.

Our results lead to interesting open problems. One such problem concerns possible de-autonomisation of maps arising from KHK discretisation 
\cite{Ramanietal1991}.
Following the results of \cite{Mase2019JIntSys} we should be
able to realise this de-autonomisation as one of the discrete
Painlev\'e equations described in \cite{Sakai2001} 
obtained by considering a surface equipped with an anti-canonical divisor 
which is blown up at 9 points in $\Pj^{2}$.
See also \cite{Joshi2019book,KajiwaraNoumiYamada2017R} for a 
comprehensive list of discrete Painlev\'e equations and the symmetry
groups of the associated elliptic surfaces on $\PcrossP$.

The surface $S$ constructed in section
\ref{sec:resolution} is not minimal, as there exists lines like
$\left\{ x\pm y=0 \right\}$ which have self-intersection $-3$.
However, finding the corresponding minimal model 
is a non-algorithmic and non-trivial task, which we will address
in future works. A final open question is whether there exist partial difference (or, lattice) equations 
which give such KHK maps as periodic or generalised reductions.

\section*{Acknoledgments}
The research reported in this paper was supported by the 
  Australian Laureate Fellowship \#FL120100094, and Discovery
  project \#DP190101838 from the Australian Research Council.

We would like to thank  Dr. Milena Radnovi\'{c} for the
interesting and fruitful discussion during the preparation of
this paper. This research was supported by an Australian Laureate Fellowship 
\# FL120100094 and grant \# DP160101728 from the Australian Research Council.

\bibliographystyle{srtnumbered}
\bibliography{bibliography}
\end{document}